\documentclass[aps,prd,superscriptaddress,12pt,showpacs,notitlepage]{revtex4}
\usepackage{graphics}
\usepackage{graphicx}
\usepackage{amsfonts}
\usepackage{amsmath}

\newcommand{\be}{\begin{equation}}
\newcommand{\ee}{\end{equation}}
\newcommand{\bea}{\begin{eqnarray}}
\newcommand{\eea}{\end{eqnarray}}
\newcommand{\pa}{\partial}
\newcommand{\bb}{\bibitem}
 
\begin{document}
\title{An algebraic construction of twin-like models}
\author{C. Adam}
\affiliation{Departamento de F\'isica de Part\'iculas, Universidad de Santiago de Compostela and Instituto Galego de F\'isica de Altas Enerxias (IGFAE) E-15782 Santiago de Compostela, Spain}
\author{J.M. Queiruga}
\affiliation{Departamento de F\'isica de Part\'iculas, Universidad de Santiago de Compostela and Instituto Galego de F\'isica de Altas Enerxias (IGFAE) E-15782 Santiago de Compostela, Spain}

\pacs{11.30.Pb, 11.27.+d}

\begin{abstract}
If the generalized dynamics of K field theories (i.e., field theories with a non-standard kinetic term) is taken into account, then the possibility of so-called twin-like models opens up, that is, of different field theories which share the same topological defect solution with the same energy density.
These twin-like models were first introduced in Phys. Rev. D{\bf 82}, 105006 (2010), Ref. \cite{trodden}, where the authors also considered possible cosmological implications and gave a geometric characterization of twin-like models. A further analysis of the twin-like models was accomplished in Phys. Rev. D{\bf 84}, 045010 (2011) , Ref. \cite{bazeia1},  with the help of the first order formalism, where also the case with gravitational self-interaction was considered. Here we show that by combining the geometric conditions of Ref. \cite{trodden} with the first order formalism of \cite{bazeia1}, one may easily derive a purely algebraic method to explicitly calculate an infinite number of twin field theories for a given theory. We determine this algebraic construction for the cases of scalar field theories, supersymmetric scalar field theories, and self-gravitating scalar fields. Further, we give several examples for each of these cases.
\end{abstract}

\maketitle 

\section{Introduction}
There exist wide classes of classical non-linear field theories which support topological defect solutions. These topological defect solutions typically have their energy densities concentrated in a certain finite region of space and are stable, where their stability is related to topological properties of the base and target spaces. Here, a nontrivial topological structure in base space (e.g., an effective compactification) is usually induced by the requirement of finite energy. These topological defects have found applications in many fields of physics, and in particular may have important applications in the field of cosmology. On the one hand, they may be relevant for structure formation in the early universe, and for its resulting evolution. Indeed, if the very early universe passed through a phase transition from a symmetric to a symmetry-breaking phase, then in the broken phase topological defects may have formed and influenced the distribution of matter and energy, see e.g. \cite{Vil}, \cite{Hind}, \cite{Battye}. 
On the other hand, there exists the idea that the whole visible universe might be just a topological defect in some higher-dimensional bulk space, the so-called brane-world scenario. The brane (i.e., our universe) in this scenario may be either strictly 3+1 dimensional ("thin brane") or have a small but nonzero extension also in the additional dimensions ("thick brane"). In the latter, thick brane case, these branes are normally  topological defects in the higher-dimensional bulk space \cite{Akama1} - \cite{Dzhu1}. In these applications, the relevant topological defects are usually solutions of some (effective or fundamental) scalar field theory, where the theory may either be of the standard type (standard kinetic term plus a potential) or of a more general type where the Lagrangian density may be a general function of the fields and their first derivatives. These generalized theories where the kinetic term does not have to be of the standard form (so-called K field theories) have already found some applications, beginning with the observation about a decade ago that they might be relevant for the solution of some problems in cosmology, like K-inflation \cite{k-infl} and K-essence \cite{k-ess}. Further applications of K fields to cosmological issues may be found, e.g., in \cite{A-H1} - \cite{liu}, whereas other, more formal or mathematical aspects of K field theories, like the existence of topological defects with compact support (so-called compactons) have been studied, e.g., in \cite{werle} -\cite{fring}. Well-posedness of the K field system and the issue of signal propagation in K field backgrounds has been investigated, e.g., in \cite{bab-muk-1} and, recently, in \cite{bergli1}.

The larger class of models allowed by generalized K field theories introduces further scales into the system under consideration via additional dimensionful couplings, therefore the resulting topological defects are, in general, quite different from their standard counterparts, see e.g. 
\cite{babichev1}, \cite{bazeia3}, \cite{comp}. Quite recently it has been found, however, that there exists the possibility that a topological defect of a non-standard K field theory perfectly mimics a defect of a standard field theory by coinciding with the standard solution both in the profile
(i.e., in the defect solution itself) and in the corresponding energy density \cite{trodden}. These coinciding solutions with their coinciding energy densities were dubbed twin or Doppelg\"anger defects in Ref. \cite{trodden}. The shape (profile) of a defect together with its energy density are the physically most relevant properties of a defect in a cosmological setting, therefore the possibility of these twins implies that, e.g., the influence of a pattern of K defects on the evolution of the universe could be mimicked by its standard twin, or vice versa. More generally, all measurable physical properties which are determined by the field profile and the energy density are indistinguishable between the K field theory and its standard twin. A more refined analysis shows, however, that there remain some differences between two twin-like models. The spectrum of linear fluctuations about the K field theory and its standard twin, for instance, are in general different \cite{trodden}, \cite{bazeia1}. The authors of \cite{trodden} discussed the example of a Dirac--Born--Infeld (DBI) type twin of a standard field theory in some detail, motivated by string theory considerations. They also gave a geometric characterization which possible twins of a standard theory have to obey and concluded from these that there exist, in general, infinitely many K field twin models for a given standard scalar field theory. The study of twin-like models was carried further in Ref. \cite{bazeia1}, where the authors employed the first order formalism in their analysis. They also considered the case with gravitational backreaction in 4+1 dimensions, where their results are of direct relevance for the brane world scenario. Further, they gave explicit examples of all cases they considered.

It is the purpose of the present acticle to derive a purely algebraic method for the construction of K field twins of a given scalar field theory which does not require knowledge of either the defect solution or its energy density. This algebraic construction may be found by combining the geometric characterization of twins of Ref. \cite{trodden} with the first order formalism of \cite{bazeia1} and allows to explicitly calculate an infinite number of twin field theories for any given scalar field theory. 

Our paper is organized as follows. In section II we briefly review the first order formalism and the geometric characterization of twins. Then we explain the algebraic construction of twin models and give several explicit examples among which the examples of Refs. \cite{trodden} and \cite{bazeia1} can be found. We also briefly discuss stability issues (energy positivity and the null energy condition (NEC)).
In Section III we repeat the same analysis for supersymmetric K field twins of supersymmetric scalar field theories. Here, one important pillar of the construction is, of course, the fact that supersymmetric K field theories exist at all, which has been demonstrated recently \cite{susy2}, \cite{bazeia2}, \cite{susy-bS} (for supersymmetric K field theories in 3+1 dimensions see \cite{ovrut1}, \cite{ovrut2}). Defects of supersymmetric theories may be of cosmological relevance if the formation of these defects occurs at time or energy scales when supersymmetry is still unbroken.
In section IV we consider the case of a self-gravitating scalar field in arbitrary dimensions, where the defects are of the wall type (i.e., still co-dimension one defects, like in the previous sections). We again derive the purely algebraic construction of K field twins of a self-gravitating standard field theory with topological defect solutions and provide several examples. In 3+1 dimensions these are just the defect solutions which are required for cosmological considerations, but now with the gravitational backreaction taken into account. In 4+1 dimensions the defects are the ones relevant for the brane world picture, where we also rederive the example already given in \cite{bazeia1}.
Section V contains a discussion of our results.

\section{Twin-like models}
\subsection{Generalized K fields and first order formalism}

The first order formalism for generalized K fields has been developed, e.g., in \cite{bazeia3}, to which we refer for a more detailed discussion. Here we just review those aspects which we shall need in the subsequent discussion.
For a general Lagrangian ${\cal L}(X,\phi)$ where $X\equiv \frac{1}{2} \pa_\mu \phi \pa^\mu \phi = \frac{1}{2}(\dot{\phi}^2 - \phi '^2)$, the energy momentum tensor reads
\be
T_{\mu\nu} = {\cal L}_{,X} \pa_\mu \phi \pa_\nu \phi - g_{\mu\nu} {\cal L}
\ee
and the Euler--Lagrange equation is
\be
\pa_\mu ( {\cal L}_{,X} \pa^\mu \phi ) - {\cal L}_{,\phi} =0
\ee
For static configurations $\phi = \phi (x)$, $\phi ' \equiv \pa_x \phi$, the nonzero components of the energy momentum tensor are
\bea \label{stat-en-de}
T_{00} &=& {\cal E} = -{\cal L} \\
T_{11}  &=& {\cal P} = {\cal L}_{,X} \phi'^2 + {\cal L}
\eea
where ${\cal E}$ is the energy density and ${\cal P}$ is the pressure. The static Euler--Lagrange equation reads
\be
({\cal L}_{,X} \phi ')' + {\cal L}_{,\phi} =0
\ee
and, after multiplication with $\phi '$, may be integrated once to give
\be
-2X{\cal L}_{,X} + {\cal L} = \phi '^2 {\cal L}_{,X} + {\cal L} \equiv {\cal P} =c
\ee
where $c$ is an integration constant. For our purposes the only acceptable value of this constant is zero for the following reason. All the models we shall consider will have one or several (constant) vacuum values $\phi = \phi_{0i}$, $i=1,...,n$, where the energy density takes its minimum value, and this minimum value is equal to zero (this may always be achieved by adding a constant to the Lagrangian). Further, static finite 
energy solutions (kinks) have to approach vacuum values for $|x|\to \infty$, which implies that for these finite energy solutions $c$ in the above equation must be zero in the same limits. But $c$ is a constant, so it is zero everywhere.  Therefore, the once integrated field equation for static fields (or zero pressure condition) in our case reads ($\phi'^2 = -2X$)
\be \label{first-order-eq}
-2X{\cal L}_{,X} + {\cal L} = 0.
\ee
Eq. (\ref{first-order-eq}) is a nonlinear first order ODE, but sometimes it is preferable to view it just as an algebraic equation for $\phi'$ with one or several ($N$) pairs of roots 
\be \label{roots}
(\phi '_i )^2 = f_i(\phi)^2 \quad \Rightarrow \quad
\phi '_i = \pm f_i (\phi) \; , \quad i=1\ldots N
\ee
as solutions. A kink solution will, in general, be the solution to one of these roots (when viewed as a first order ODE), or it may even be the result of joining different solutions in a smooth way.   

It is one of the virtues of the first order formalism that the knowledge of the roots  (\ref{roots}) together with the asymptotic values (i.e., vacuum values) $\phi_\pm \equiv \phi_k (\pm \infty) $ of the kink solution $\phi_k (x)$ is sufficient for the calculation of the kink energy, i.e., one does not need the explicit solution $\phi_k (x)$. The important point is that with the help of the corresponding root, the energy density of a kink may be viewed as a function of either only $\phi$ or only $\phi '$.   This allows one to separate a factor $\phi '$ from the energy density which, together with the base space differential $dx$ in the energy functional, may be traded for a target space differential according to $d\phi = dx \phi '$. The remainder must, of course, be interpreted as a function of $\phi$ only. Explicitly, the energy reads
\be
E = \int_{-\infty}^\infty dx {\cal E} 
\equiv  \int_{-\infty}^\infty dx \phi ' W_{,\phi} = \int_{\phi (-\infty)}^{\phi (\infty)} d\phi W_{,\phi} = W(\phi (\infty)) - W(\phi (-\infty))
\ee
where $W_{,\phi}$ (and its $\phi$ integral $W(\phi)$) must be interpreted as a function of $\phi$ only, which results upon replacing $\phi '$ by its corresponding root $f_i (\phi)$ in the above expression.

For theories with a standard kinetic term $X$ and a potential $V(\phi)$,
\be
{\cal L}_{s} = X-V,
\ee
the integrated static field equation simply is
\be \label{s-first-order-eq}
-X-V=0 \quad \Rightarrow \quad \phi'^2 = 2V
\ee
with the two roots $\phi ' = \pm \sqrt{2V}$. If the potential $V$ has at least two vacua (which we assume from now on), then there will exist, in general, finite energy solutions of Eq. (\ref{s-first-order-eq}) which interpolate between different vacua (kinks), and the two roots correspond to kink and antikink, respectively. The static energy density for the standard theory is
\be
{\cal E}_{s} = -X+V = \frac{1}{2}\phi '^2 + V
\ee
and for a kink solution it may be written as
\be
{\cal E}_{s}|_{\phi_k} = (-X+V)|_{\phi_k} = 2V({\phi_k}) = -2X|_{\phi_k}
\ee
where $\phi_k (x)$ is the kink solution under consideration, and the notation $|_{\phi_k}$ means that the expression (in general, a function of $\phi$ and $\phi '$), is evaluated at the kink solution $\phi =\phi_k (x)$.
Finally, the energy of a kink in this standard case simply is
\be
E = \int_{-\infty}^\infty dx \phi '^2 = \int_{\phi (-\infty)}^{\phi (\infty)} d\phi \phi ' =  \int_{\phi (-\infty)}^{\phi (\infty)} d\phi (\pm \sqrt{2V})
= \pm [W_V(\phi (\infty)) - W_V(\phi (-\infty))]
\ee
where 
\be
W_{V,\phi} = \sqrt{2V}
\ee
and the explicit expression for $W_V$ depends, of course, on $V$. The two signs correspond to kink and antikink, respectively.

\subsection{Twin or doppelgaenger defects}

In \cite{trodden} the authors observed the possibility of twin-like models within the class of generalized K field theories, that is, of field theories which share the same kink solution with the same energy density with a given standard field theory. They discussed a Dirac-Born-Infeld (DBI) like example in some detail where, however, the DBI term is multiplied by a target space geometric factor, because a pure DBI theory cannot be the twin of a standard field theory. Then they derived a necessary and sufficient geometrical condition which a second field theory ${\cal L}_2$ has to obey in order to be the twin of a given field theory ${\cal L}_1$. From their geometric description they already concluded that there exist, in principle, infinitely many twin theories for a given standard scalar field theory. We shall review this geometric construction in a first step, because we will find that combining it with the first order formalism provides us with a simple and purely algebraic method to explicitly calculate an infinite number of twin models for any given field theory. The authors of \cite{trodden} demonstrated that if the theory ${\cal L}_1$ has a kink solution $\phi_k (x)$ with energy density ${\cal E}_k(x)$, then a necessary and sufficient condition for a second theory ${\cal L}_2$ to have the same kink solution with the same energy density is that both ${\cal L}$ and ${\cal L}_{,X}$ agree when evaluated for the kink solution, that is,
\bea \label{cond1}
{\cal L}_1 |_{\phi_k} &=& {\cal L}_{2}|_{\phi_k} \\
 {\cal L}_{1,X} |_{\phi_k} &=& {\cal L}_{2,X} |_{\phi_k}. \label{cond2}
\eea 
Obviously, the first condition implies that the energy densities are equal, see Eq. (\ref{stat-en-de}). Further, the first order equation Eq. (\ref{first-order-eq}) holds for ${\cal L}_1$ by assumption, then the two conditions (\ref{cond1}) and (\ref{cond2}) imply that Eq. (\ref{first-order-eq}) is an identity for ${\cal L}_2$. It follows that the two conditions (\ref{cond1}) and (\ref{cond2}) are sufficient for ${\cal L}_2$ to be a twin of ${\cal L}_1$. That the two conditions are necessary follows easily from the fact that the two equations (\ref{stat-en-de}) and (\ref{first-order-eq}) are linear in ${\cal L}$ and ${\cal L}_X$. 

From what has been said above, it might appear that for the explicit construction of a twin model ${\cal L}_2$ for a given theory ${\cal L}_1$ it is necessary to know an explicit kink solution $\phi_k$ of the theory ${\cal L}_1$, and to use this kink in the evaluation of possible twin models ${\cal L}_2$, which would render calculations rather cumbersome. But this is, in fact, not true. The important point is that the lagrangian densities are functions of the target space variables $\phi$ and $\phi '$ only, therefore it is sufficient to implement the root $\phi ' = \pm f_i (\phi)$ which leads to the kink (or antikink) solution under consideration. Further, we shall use the fact that all lagrangians we consider depend on $\phi '$ only via $X=-\frac{1}{2}\phi '^2$ (for static configurations), so that the above conditions transform into
\bea \label{cond1a}
{\cal L}_1 |_{2X=-f^2_i} &=& {\cal L}_{2}|_{2X=-f^2_i} \\
 {\cal L}_{1,X} |_{2X=-f^2_i} &=& {\cal L}_{2,X} |_{2X=-f^2_i} \label{cond2a}
\eea 
where $f_i (\phi)$ is a known root (\ref{roots}) of the theory ${\cal L}_1$ leading to a kink solution. The above conditions are purely algebraic conditions in the target space variables $\phi$ and $X$ and do not involve the base space variable $x$ or explicit knowledge of a kink solution $\phi_k (x)$ at all. 

Up to now we allowed for completely general lagrangians ${\cal L}_1$ and ${\cal L}_2$ to emphasize the general character of the procedure. Now, however, we will concentrate on the case of a standard lagrangian ${\cal L}_1 = {\cal L}_s =X-V$ for concreteness, so the problem consists in finding possible twins to standard scalar field theories. Here $V(\phi)$ is a positive semi-definite potential with at least two vacua (zeros) such that kink solutions exist. The two roots for kink and antikink may be combined into $X=-V$, and the above conditions read (we write ${\cal L}$ for ${\cal L}_2$) 
\bea \label{cond1b}
{\cal L}|_{X=-V} &=& -2V \\
 {\cal L}_{,X} |_{X=-V} &=& 1. \label{cond2b}
\eea 
Again, these two conditions are purely algebraic and allow an easy calculation of twin models, as we shall see in the next section.

\subsection{Examples of twin models}

As a first class of twin models let us consider the class of Lagrangians
\be \label{power-ex}
{\cal L}= \sum_{k=1}^K f_k (\phi) X^k - U(\phi)
\ee
where the kinetic terms $X^k$ are multiplied by functions of $\phi$ in a sigma-model like fashion. Here, the condition
\be \label{cond2c}
{\cal L}_{,X}|_{X=-V} =\sum_{k=1}^K kf_k(\phi) (-V)^{k-1} \equiv 1
\ee
imposes one condition on the functions $f_k (\phi)$. One may, for instance, choose arbitrary $f_k$ for $k\ge 2$, then the above condition determines $f_1$ in terms of the remaining $f_k$ and $V$. We remark that it is not possible to choose all $f_k$ constant, but if at least one $f_k$ has a nontrivial $\phi$ dependence then the above condition can always be fulfilled. The second condition 
\be \label{cond1c}
{\cal L}|_{X=-V} = \sum_{k=1}^K f_k (\phi ) (-V)^k - U(\phi) \equiv -2V(\phi) ,
\ee
in turn, determines $U(\phi)$ in terms of the $f_k$ and $V$. 

One question to ask is whether the resulting twin models constitute viable field theories on their own, that is, whether they obey certain stability requirements like energy positivity or the null energy condition (NEC). Here we shall mainly be concerned with the NEC, because $i)$ it is deemed sufficient for stability, $ii)$ it is weaker than the condition of positivity of the energy density and $iii)$ it is easier to implement for the class of models we study in this paper. The NEC in general is the condition that
\be
n^\mu n^\nu T_{\mu\nu} \ge 0
\ee
where $T_{\mu\nu}$ is the energy-momentum tensor and $n^\mu$ is an arbitrary null vector. For the class of models ${\cal L}(X,\phi)$ the NEC simply reads
\be
{\cal L}_{,X} \ge 0.
\ee    
It is, in general, not completely trivial to reconcile the NEC with the two twin conditions (\ref{cond2c}) and (\ref{cond1c}), but it is easy to find certain special classes of models where the NEC holds by construction. 

A first class of models which obeys both the NEC and the condition (\ref{cond2c}) by construction is given by field theories which obey
\be
{\cal L}_{,X} = K f(\phi) (X+V)^{K-1} +1
\ee
where $f$ is an arbitrary, positive semi-definite function $f(\phi) \ge 0$ and $K$ is an odd integer. The resulting Lagrangian (i.e., $X$ integral) is
\be \label{lag-tilde-U}
{\cal L} = f(\phi) (X+V)^K + X - \tilde U(\phi)
\ee
(where the integration "constant" $\tilde U(\phi)$ is an arbitrary function of $\phi$),
and the second twin condition (\ref{cond1c}) requires $\tilde U =V$ such that the class of twin Lagrangians reads
\be
{\cal L} = f(\phi) (X+V)^K + X - V \; , \quad K=3,5,\ldots
\ee
As a concrete example, we may e.g. choose $f=1$ and $K=3$ which results in the Lagrangian
\be
{\cal L}= \frac{1}{3}X^3 + V X^2 + (V^2 +1) X + \frac{1}{3}V^3 - V
\ee
which shares both the kink solution $\phi' = \pm \sqrt{2V}$ and the corresponding energy density with the standard scalar model ${\cal L}_s = X-V$.  
A second class of twin models obeying the NEC may be constructed from the equation
\be
{\cal L}_{,X} = f^{1-K} (X+V+f)^{K-1} 
\ee
(where $f=f(\phi)\ge 0$, and $K$ is an odd integer)
with Lagrangian
\be
{\cal L} = \frac{f^{1-K}}{K} (X+V+f)^K -\tilde U.
\ee
Here the second twin condition leads to $\tilde U = 2V+(f/K)$ and, therefore, to the Lagrangian
\be
 {\cal L} = \frac{f^{1-K}}{K} (X+V+f)^K -2V - \frac{f}{K} .
 \ee
 
Next, let us describe another class of examples of twin models, different from the power expansion in $X$ of Eq. (\ref{power-ex}). We start from the ansatz
\be
{\cal L}= f(\phi) g(X) - U(\phi)
\ee
and calculate
\be
{\cal L}_{,X} = f(\phi) g' (X)
\ee
and the NEC leads to the conditions
\be
f\ge 0 \; , \quad g' \ge 0.
\ee
Further, the two twin conditions lead to $f(\phi)=(g' (-V))^{-1}$ and $U=2V + (g(-V)/g'(-V))$ and, therefore, to the Lagrangian
\be
{\cal L}= \frac{g(X)}{g'(-V)} -\frac{g(-V)}{g'(-V)} -2V .
\ee
Among this class we may easily recover the DBI type example originally presented and discussed in \cite{trodden}. Indeed, choosing for the kinetic function $g(X)$ the DBI type expression
\be
g(X) = -\sqrt{1-2X}
\ee
we calculate 
\be
g' (X) = \frac{1}{\sqrt{1-2X}} \; ,\quad f(\phi )=\sqrt{1+2V} \; ,\quad U=2V-(1+2V)=-1
\ee
and the resulting Lagrangian is
\be
{\cal L}= -\sqrt{1+2V}\sqrt{1-2X} +1.
\ee
It is obvious from the derivation that the nontrivial target space geometry factor $f(\phi)=\sqrt{1+2V}$ is necessary for this DBI type action to be the twin  of a standard scalar field theory, as announced above.

\section{Supersymmetric twin models}

To begin with, let us remind that a standard scalar field theory ${\cal L}_s = X-V$ with a positive semi-definite potential $V\ge 0$ may always be viewed as the purely bosonic sector of a supersymmetric scalar field theory. Indeed, before the elimination of the auxiliary field $F$ the bosonic sector of the supersymmetric standard scalar field theory reads
\be
{\cal L}_s = \frac{1}{2} (\pa_\mu \phi \pa^\mu \phi + F^2) - F P_s' (\phi)
\ee
where $P_s (\phi)$ is the prepotential (also sometimes called superpotential) of the standard SUSY scalar field theory. Elimination of the auxiliary field $F$ with the help of its algebraic field equation $F=P_s'$ leads to the lagrangian
\be
{\cal L}_s = \frac{1}{2} \pa_\mu \phi \pa^\mu \phi -\frac{1}{2} P_s'^2
\ee
which is just the standard scalar field lagrangian with the identification
\be
V= \frac{1}{2} P_s'^2 \ge 0.
\ee
This observation leads to the obvious question whether there exist supersymmetric K field theory twins for the supersymmetric standard field theories.  For this purpose, in a first instance we have to know whether there exist supersymmetric scalar K field theories at all. The answer is that these supersymmetric K theories do exist. Some classes of examples have been introduced and studied in \cite{susy2}, \cite{bazeia2} (these theories exist both in 1+1 and in 2+1 dimensional Minkowski space, due to the similar spin structure in the two spaces), and we shall use some of these examples for the construction of our supersymmetric K field twins. 
In \cite{susy2} a class of supersymmetric models was introduced such that their purely bosonic sector before the elimination of the auxiliary field reads
\be
{\cal L}^{(\alpha ,P)} 
= \sum_{k=1}^N \alpha_k (\phi) [ (\partial^\mu\phi\partial_\mu\phi)^k + (-1)^{k-1}F^{2k}] - P'(\phi) F.
\ee
Next, the auxiliary field $F$ should be eliminated via its algebraic field equation
\be \label{F-eq}
\sum_{k=1}^N (-1)^{k-1} 2k \alpha_k F^{2k-1} - P' (\phi )=0
\ee
which in general is, however, a rather complicated algebraic equation for $F$. As no assumption was made yet about the functional dependence of $P$, this equation may be understood in a second, equivalent way: one assumes that $F$ is an arbitrary given function of $\phi$, which in turn determines the prepotential $P(\phi)$.  This second way of interpreting Eq. (\ref{F-eq}) is more useful for our purposes. Eliminating the resulting $P'(\phi)$ we arrive at the Lagrangian density
\bea \label{L-b-F}
{\cal L}^{(\alpha ,F)}
&=& \sum_{k=1}^N \alpha_k (\phi) [ (\partial^\mu\phi\partial_\mu\phi)^k - (-1)^{k-1}(2k-1) F^{2k}] 
\eea
where now $F=F(\phi)$ is a given function of $\phi$ which may be chosen freely depending on the theory or physical problem under consideration. 
This class of lagrangians is exactly of the type (\ref{power-ex}), therefore the conditions for being the twin of a standard theory are exactly analogous to the conditions (\ref{cond1c}) and (\ref{cond2c}). 
The restrictions implied by supersymmetry (i.e., the requirement to express the "potential function " $U(\phi)$ in Eq. (\ref{power-ex})  in terms of $F(\phi)$), nevertheless, impose some additional restrictions, as we want to show now. Indeed, the second twin condition $L_{,X}\vert_{2X=-F_s^2}=1$ leads to
\be \label{susy-twin-2}
\sum 2k \alpha_k (-F_s^2)^{k-1} =1,
\ee
where we introduced the function $F_s(\phi)$ of the standard SUSY theory, i.e., the auxiliary field $F$ of the standard theory evaluated at its field equation via 
\be \label{Fs}
2V(\phi) \equiv P_s{}'^2 (\phi) \equiv F_s^2 (\phi)
\ee 
for convenience.
The first twin condition $ L\vert_{2X=-F_s^2}=-F_s^2$ then leads to
\bea \nonumber 
L\vert_{2X=-F_s^2} &=& \sum_k \alpha_k ((-F_s^2)^k - (-1)^{k-1} (2k-1) F^{2k}) \\
&=& \sum_k \alpha_k ((-F_s^2)^k - (-F^2)^k - 2kF^2 (-F^2)^{k-1}) \nonumber \\
&\equiv & -F_s^2 \nonumber
\eea
where we used (\ref{susy-twin-2}) in the last step. This condition is solved by 
\be \label{F=Fs}
F=\pm F_s.
\ee 
 As we shall see in a concrete example below, this is typically the only acceptable solution, therefore supersymmetry seems to imply the additional relation $F=F_s$ for the algebraic solutions of the auxiliary fields of standard and K field twin theories.

Again, the NEC is not automatic in these theories, but a more specific class of examples which obeys the NEC by construction may be given, analogous to the last subsection. Concretely, we give some examples starting from the $X$ derivative
\be
{\cal L}_{,X} = 2K (2X + 2V)^{K-1} +1
\ee
(we write $2X$ instead of $X$ in order to be as close as possible to the notation used in Ref. \cite{susy2} and in Eq. (\ref{L-b-F}); $K$ is an odd integer). 
The resulting Lagrangian is 
\be
{\cal L} = (2X+2V)^K + X - \tilde U (\phi )
\ee
and obeys the NEC and the twin condition ${\cal L}_{,X}|_{X=-V}=1$ by construction. 

For a more concrete example, let us now assume $K=3$ which leads to the Lagrangian
\be
{\cal L} = (2X)^3 + 6V (2X)^2 + (12 V^2 + \frac{1}{2}) (2X) + 8V^3 - \tilde U
\ee
and therefore to
\be
\alpha_3 = 1 \; , \quad \alpha_2 = 6V \; , \quad \alpha_1 = 12 V^2 + \frac{1}{2}
\ee
and to the Lagrangian
\be
{\cal L}= (2X)^3 - 5F^6 + 6 V ((2X)^2 + 3F^4 ) + (12 V^2 + \frac{1}{2} )(2X-F^2 )
\ee
 which explicitly is of the form (\ref{L-b-F}) (we replaced the arbitrary integration "constant" $\tilde U (\phi )$ by the required $F$ terms). 
Now the second twin condition ${\cal L}|_{X=-V} = -2V$ leads to
\be
5F^6 - 18 V F^4 + (12 V^2 + \frac{1}{2}) F^2 + 8 V^3 - V =0
\ee
which may be viewed as a third order algebraic equation for $F^2$. The only acceptable (i.e., real and positive) solution is 
\be \label{F-2=2V}
F^2 =2V \equiv F_s^2
\ee  
and leads to the Lagrangian 
\be
{\cal L}= (2X)^3 - 40 V^3 +6V((2X)^2 + 12 V^2 ) + (12 V^2 + \frac{1}{2}) (2X-2V)
\ee
which is the desired supersymmetric twin  of the standard Lagrangian. As already remarked for the more general class of examples above, 
it holds that also the (algebraic) field equations for the auxiliary fields coincide, see Eq. (\ref{F=Fs}). This equality is {\em not} a further condition, but a consequence of the twin conditions and supersymmetry.

Another class of supersymmetric theories has the following purely bosonic sector (before the elimination of the auxiliary field $F$) \cite{bazeia2}
\be
{\cal L}= g(\phi )f(X) (F^2 + 2X^2 ) - P'(\phi) F
\ee 
where $f$ and $g$ are arbitrary, fixed functions of their arguments and for the moment we just assume $g\ge 0$. 
The algebraic field equation for the auxiliary field $F$ has the solution
\be \label{F-eq2}
F = \frac{P'}{2gf}
\ee
and leads to the Lagrangian
\be
{\cal L}= 2Xgf -\frac{P'^2}{4gf} \equiv 2g\left( Xf - \frac{h}{f}\right)
\ee
where
\be
h(\phi) \equiv \frac{P'^2}{8g^2} .
\ee 
The $X$ derivative of this lagrangian is
\be
{\cal L}_{,X} = 2g\left( f+Xf_{,X} + h \frac{f_{,X}}{f^2} \right) .
\ee
A sufficient condition for the NEC consists in the following inequalities
\be
f+Xf_{,X} \ge 0 \; ,\quad f_{,X} \ge 0
\ee
but we have not been able to find a function $f$ which obeys these inequalities. There exists, however, another possiblity to obey the NEC, and for this possibility we found solutions. Concretely, assume that
\be
f+Xf_{,X} \ge 1
\ee
and that further
\be
\left| \frac{f_{,X}}{f^2} \right| \le 1
\ee
and 
\be
h\le 1
\ee
then the NEC holds. A specific function $f$ obeying these conditions is
\be
f=1+X^2
\ee
which indeed leads to 
\be 
f+Xf_{,X} = 1+3X^2 \ge 1 \; ,\quad \left| \frac{f_{,X}}{f^2} \right| = \frac{2|X|}{(1+ X^2)^2} \le 1.
\ee
We will study this explicit example in what follows. We remark that, as we shall see,  the condition $h\le 1$ leads to restrictions on possible potentials $V$, so if we insist on the NEC we may construct twins of the type considered here only for standard theories with certain potentials. 
For the specific choice $f=1+X^2$ the lagrangian and its $X$ derivative read
\bea
{\cal L} &=& 2g\left( X+X^3 -\frac{h}{1+X^2}\right) \\
{\cal L}_{,X} &=& 2g \left(1+3X^2 +h\frac{2X}{(1+X^2)^2}\right) .
\eea
The twin condition ${\cal L}_{,X}\vert_{X=-V} =1$ leads to the equation
\be \label{h-eq}
h=\frac{(1-V^2)^2}{2V}\left( 1+3V^2 -\frac{1}{2g} \right)
\ee
and the second twin condition ${\cal L}\vert_{X=-V}=-2V$ leads, together with Eq.  (\ref{h-eq}), to the solution
\be
\frac{1}{2g} = 1+X^2 
\ee
which, in turn, leads to
\be
h=V(1+V^2)^2 .
\ee
Now, the NEC requires $h\le 1$ which obviously restricts possible potentials $V$. An example of a potential which is compatible with this condition is 
\be
V=\frac{1}{2}\frac{(1-\phi^2)^2}{(1+\phi^2)^2}
\ee
as may be checked easily. Further, this potential has the same vacuum structure as the standard $\phi^4$ potential $V=(1/2)(1-\phi^2)^2$, so it will lead to similar kink solutions. 

We want to end this section with the remark that the auxiliary field $F$, when evaluated at the kink equation $X=-V$, again coincides with the auxiliary field of the standard supersymmetric theory $F_s^2 = 2V$. Indeed, from Eq. (\ref{F-eq2}) we infer that
\be
F^2 = \frac{P'^2}{4g^2f^2} = \frac{2h}{f^2} 
\ee
which depends both on $\phi$ and on $X$. But evaluating it for the kink equation leads to $f\vert_{X=-V} = 1+V^2$,
which together with the solution $h=V(1+V^2)^2$ just leads to
\be
F^2 \vert_{X=-V} = \frac{2V(1+V^2)^2}{(1+V^2)^2} =2V \equiv F_s^2
\ee
which is, again, identical to the field equation of the auxiliary field for the standard supersymmetric scalar field theory. 

\section{Self-gravitating twins}

Here we want to study the existence of twins of the standard scalar field theory fully coupled to gravity, that is, K field theories which give rise to exactly the same defect solution, energy density, and induced metric than the standard scalar field theory with self-gravitation fully taken into account. We shall find that the situation is completely equivalent to the Minkowski space case in that, again, there exist two purely algebraic "twin conditions" which allow to calculate twins of self-gravitating standard scalar field theories. The only differences will be that $i)$ the "on-shell" condition for a defect is no longer $X=-V$ but, instead, $X=-(1/2)W_{,\phi}{}^2$, where the relation between $W$ and $V$ is slightly more complicated than in the flat space case; and $ii)$ the "on-shell" value which the Lagrangian has to take will be different, as well, i.e., ${\cal L}\vert_{X=-(1/2)W_{,\phi}{}^2}=-W_{,\phi}{}^2 + c_d W^2$ instead of ${\cal L}\vert_{X=-V}=-2V$ (here $c_d$ is a numerical coefficient which depends on the dimension $d$ of space-time; in principle, it also depends on the gravitational constant $\kappa$ and vanishes in the limit $\kappa \to 0$, but we shall choose units such that $\kappa =1$ in the following).   

Before starting the detailed calculations, some remarks are in order. The topological defect solution in flat Minkowski space may be either viewed as a kink solution in 1+1 dimensions or as a co-dimension one domain wall solution in higher dimensions. Both the defect solution and its energy density per length unit in the direction perpendicular to the wall do not depend on the dimension. This is no longer true once the gravitational self-interaction is taken into account. In 1+1 dimensions there is no gravitational interaction, because the Einstein tensor is identically zero, and for higher dimensions $d>2$ the Einstein equations depend on the dimension $d$, therefore also the self-gravitating defect solutions will depend on $d$. Here we shall discuss  the case for general $d$, but probably the two cases $d=4$ and $d=5$ are the most interesting ones. $d=4$  is the dimension of our universe, at least at a macroscopic scale, so the resulting defects of the standard theory and its twins may be viewed just as domain walls in the universe. The case $d=5$ is especially interesting in relation to the braneworld scenario, where our universe is identified with the domain wall, and the direction perpendicular to the domain wall is identified with a fifth direction or coordinate which is invisible due to the resulting warped geometry in five dimensions, which essentially confines all physics to the three dimensional domain wall or brane (four dimensional brane world hypersurface). As already stated, the $d=5$ case was studied in \cite{bazeia1}, and we shall build on the results of that paper.  
Another remark concerns the possibility in flat space to express the linear energy density of a defect solution with the help of the integrating function $W$ as
${\cal E} = \phi ' W_{,\phi}$. Using the static field equation for a defect (\ref{first-order-eq}), this relation may be re-expressed like
\be
{\cal E} = -{\cal L} = \phi '^2 {\cal L}_{,X} = \phi ' W_{,\phi} \quad \Rightarrow \quad \phi ' {\cal L}_{,X} = W_{,\phi}
\ee
and this last form is the most useful one for our purposes, because it may be generalized directly to the case with gravity, as we shall see below.

If the Einstein--Hilbert action is normalized as
\be
S_{\rm EH} = \frac{1}{\kappa} \int d^d x \sqrt{|{\rm g}|}{\cal R}
\ee
(where ${\rm g}$ is the determinant of the metric ${\rm g}_{MN}$, $M,N = 0, \ldots ,d-1$, ${\cal R}$ is the curvature scalar, and $\kappa =4\pi G$ where $G$ is Newton's constant), then the Einstein equation is 
\be
G_{MN} = 2\kappa T_{MN}
\ee
where 
\be
T_{MN} = \nabla_M \phi \nabla_N \phi {\cal L}_{,X}- {\rm g}_{MN} {\cal L}
\ee
is the energy-momentum tensor. Further, $\nabla_M$ is the covariant derivative, and 
\be
X\equiv \frac{1}{2}\nabla_M \phi \nabla^M \phi .
\ee 
We shall choose length units such that $\kappa =1$, therefore the Einstein equation we use reads
\be
G_{MN} = 2 T_{MN} .
\ee
For the self-gravitating defect solution we use the ansatz for the metric
\be
ds^2 = e^{2A(y)}\eta_{\mu\nu} dx^\mu dx^\nu - dy^2
\ee
where $x^M = (x^\mu ,y)$, $y$ is the coordinate for the direction perpendicular to the domain wall (or brane), and $\eta_{\mu\nu} = {\rm diag}(+,-\ldots ,-)$ is the Minkowski metric in $d-1$ dimensions. Further, we assume that $\phi =\phi (y)$ only depends on the $y$ coordinate. With this ansatz, the expression for $X$ reduces to the same expression like in the flat space case, $X=-(1/2)(\pa_y \phi )^2 \equiv -(1/2)\phi '^2$.

The Einstein equations for this ansatz reduce to two independent equations for $A(y)$ and $\phi (y)$, and they depend on the dimensions $d$ of space-time. Explicitly, they read
\bea
 \frac{(d-1)(d-2)}{2}A'^2 + (d-2)A'' &=& 2{\cal L} \\
\frac{(d-1)(d-2)}{2}A'^2 &=& -4X{\cal L}_{,X} + 2{\cal L} 
\eea
which may be resolved for $A'^2$ and $A''$,
\bea \label{d=4-eq1}
 \quad A'' &=& \frac{4}{d-2}X{\cal L}_{,X} \\
A'^2 &=& \frac{4}{(d-1)(d-2)}({\cal L} - 2X{\cal L}_{,X}). \label{d=4-eq2}
\eea
The field equation for the scalar field $\phi$ is not an independent equation, but rather a consequence of the Einstein equations therefore we do not display it here. 
The first order formalism for static domain walls now consists in introducing an integrating function or superpotential $W=W(\phi)$ proportional to $-A'$
\cite{sken1}, \cite{dewolfe1}, \cite{FNSS},
 \cite{bazeia3}, \cite{bazeia1}. 
The right choice is
\be
 A' = -\frac{2}{d-2}W(\phi) ,
\ee
and inserting it into Eq. (\ref{d=4-eq1}) leads to
\be
A'' =-\frac{2}{d-2}W_{,\phi} \phi ' = \frac{4}{d-2}(-\frac{1}{2}\phi '^2) {\cal L}_{,X} \quad \Rightarrow \quad W_{,\phi} = \phi ' {\cal L}_{,X}
\ee
exactly as in the flat space case. In order to find the twin conditions which twin models of the standard Lagrangian ${\cal L}_s = X-V$ should obey, we first have to solve the Einstein equations for the standard Lagrangian. Obviously, the first integral for the standard Lagrangian is 
\be
\phi ' =W_{,\phi} \quad \Rightarrow \quad X=-\frac{1}{2}W_{,\phi}{}^2
\ee
just like in the flat space case. This implies that the first twin condition for a K field lagrangian is just ${\cal L}_{,X}\vert_{X=-(1/2)W_{,\phi}{}^2}=1$, in close analogy to the flat space case (although it is no longer true that $(1/2)W_{,\phi}{}^2=V$, as we shall see in a moment). In order to find the relation between $V$ and $W$ we just insert the standard Lagrangian, the ansatz for $A'$ and the first integral for the standard Lagrangian into the second Einstein equation (\ref{d=4-eq2}) and find
\be
 A'^2 \equiv \frac{4}{(d-2)^2}W^2 = \frac{4}{(d-1)(d-2)}(X-V-2X) =\frac{4}{(d-1)(d-2)}(\frac{1}{2}W_{,\phi}{}^2 -V)  
\ee
\be 
\Rightarrow \quad V= \frac{1}{2}W_{,\phi}{}^2 -\frac{d-1}{d-2} W^2 . \label{d=4-pot-eq}
\ee
We remark that the first, $W_{,\phi}$  term is exactly like in the flat space case, whereas the second, $W$ term is the correction due to gravity and depends on the dimension $d$. 
Inserting this result back into ${\cal L}_s$ leads to
\be
{\cal L}_s \vert_{X=-(1/2)W_{,\phi}{}^2} = -W_{,\phi}{}^2 +\frac{d-1}{d-2}W^2
\ee
and the resulting twin conditions for a general Lagrangian ${\cal L}$ to be the twin of a standard Lagrangian ${\cal L}_s =X-V$ are therefore
\bea \label{grav-twin1}
{\cal L}_{,X}\vert_{X=-(1/2)W_{,\phi}{}^2} &=& 1 \\ \label{grav-twin2}
{\cal L}\vert_{X=-(1/2)W_{,\phi}{}^2} &=& -W_{,\phi}{}^2 +\frac{d-1}{d-2}W^2 
\eea
where the relation between the integrating function $W$ and the potential $V$ is given in (\ref{d=4-pot-eq}). These relations are, again, purely algebraic and do not require the explicit knowledge of a defect solution. 

We remark that solving Eq. (\ref{d=4-pot-eq}) for a given potential $V$ is, in general, quite difficult. It is simpler to choose an integrating function (or superpotential) $W$ and determine the resulting potential $V$. In addition, by choosing an adequate $W$, it is also easy to assure that the simple equation $\phi '= \pm W_{,\phi}$ does support topological defect solutions. 

Finally, let us present some explicit examples. As a first example, we choose the Lagrangian (\ref{lag-tilde-U}) of Section 2, but with $V$ replaced by $\frac{1}{2}W_{,\phi}{}^2$ (where $f(\phi)$ is an arbitrary, nonnegative function),
\be 
{\cal L} = f(\phi) (X+\frac{1}{2}W_{,\phi}{}^2 )^K + X - \tilde U(\phi) \nonumber
\ee
which, by construction, obeys both the NEC and the twin condition (\ref{grav-twin1}). The second twin condition (\ref{grav-twin2}) determines $\tilde U$ to be
\be
\tilde U = \frac{1}{2}W_{,\phi}{}^2 -\frac{d-1}{d-2} W^2 \equiv V,
\ee
just like in the case without gravity.

For a second class of examples, we use the ansatz (as in Section 2)
\be
{\cal L}= f(\phi) g(X) - U(\phi).
\ee
The first twin condition (\ref{grav-twin1}) leads to
\be
f(\phi )= \frac{1}{g' (-(1/2)W_{,\phi}{}^2)}
\ee
and the second twin condition (\ref{grav-twin2}) results in
\be
U(\phi )= W_{,\phi}{}^2 + \frac{g(-(1/2)W_{,\phi}{}^2)}{g'(-(1/2)W_{,\phi}{}^2)} -\frac{d-1}{d-2}W^2
\ee
For the specific, DBI type example $g(X)=-\sqrt{1-2X}$ we, therefore, get the Lagrangian
\be
{\cal L} = -\sqrt{1+W_{,\phi}{}^2}\sqrt{1-2X} +1+\frac{d-1}{d-2}W^2
\ee
which, for $d=5$,  precisely coincides with the example presented in \cite{bazeia1}. We remark that for the case with selfgravitation
the authors of \cite{bazeia1} use the definition $W_{,\phi} =\frac{1}{2}\phi ' {\cal L}_{,X}$ for the integrating function (or superpotential) $W$, which differs by a factor two from the definition $W_{,\phi} =\phi ' {\cal L}_{,X}$ employed here (but also in Ref. \cite{bazeia1} for the case without gravity).

\section{Discussion}

In this article, we have derived a simple and purely algebraic method for the construction of K field twins of a standard scalar field theory, that is, of K field models which share the same topological defect with the same energy density with a given standard scalar field theory. This method may be derived for the cases of non-supersymmetric field theories, supersymmetric field theories and for self-gravitating fields. Further, we gave several examples for all these cases. The interest of these twin models lies in the fact that the field profile together with the energy density are the most relevant physical data of a defect which makes the twins almost indistinguishable from their standard counterparts in many situations. A pattern of defects in the very early universe will look the same irrespective of whether it is formed by defects of a standard theory or of a K field twin. 
The spectrum of linear fluctuations, on the other hand, is in general different between the standard theory and its twins \cite{trodden}, \cite{bazeia1}, so small differences will set in once dynamics (i.e., time dependence) is taken into account. A similar question is related to the behaviour of additional matter fields (e.g., fermion fields) coupled to twin defects. In the non-supersymmetric case there exist different possibilities to couple fermions to each field theory, therefore general statements cannot be made. The situation is different, however, for supersymmetric (SUSY) K field twins of standard SUSY scalar field theories. Here, the first important piece of information is, of course, the existence of SUSY K field theories \cite{susy2} - \cite{ovrut2}. We want to point out again that defects of supersymmetric theories are the relevant ones to study if the symmetry breaking and the subsequent defect formation in the early universe occur at an energy scale where supersymmetry is still intact (e.g., at the end of inflation).
For supersymmetric twins of the standard supersymmetric scalar field theories, it is interesting to observe that these SUSY twin models not only share the defect solution and its energy density with the standard theory. Also the (algebraic) field equation for the auxiliary field in the kink background is identical for the standard theory and the twin. Another interesting problem of these SUSY theories concerns, of course, the inclusion of fermions which we have set equal to zero in our discussion. In general, the fermionic sectors of the standard theory and the twin will be different, like the bosonic sectors. Standard and twin theory will, however, share some common features in the fermionic sector, too.   They will, e.g., share the same fermionic zero mode in the background of the same kink solution. This is a consequence of translational invariance, on the one hand, which implies that both theories in the kink background have the same bosonic zero mode (or Goldstone mode) equal to the derivative of the kink. The second ingredient is, of course, supersymmetry, which guarantees that each bosonic Goldstone mode is paired by a fermionic zero mode which is, again, equal to the derivative of the kink field. 

A final issue is the existence of twin models when the gravitational backreaction is taken into account, i.e., of twin defects sharing the same field profile,  energy density and induced metric. Already for defect structures in cosmology (i.e., in the early universe) the full self-gravitating case should, in principle, be taken into account, although in many circumstances a Minkowski space calculation is sufficient. In the brane world scenario taking into account the full self-gravitating solution is mandatory. We found that, again, there exists a simple algebraic method to calculate infinitely many K field twins of a standard self-gravitating scalar field theory and gave several examples. We emphasize that for the 4+1 dimensional case relevant for the brane world scenario an example of a self-gravitating twin has already been given in \cite{bazeia1} with the help of the first order formalism.

It was the main aim of the present article to shed more light on the existence of K field twin defects and the mathematical structures behind them, on the one hand, and to provide a simple calculational tool for the construction and study of twin-like models, on the other hand. 
We want to point out that, whenever K field theories cannot be excluded on purely theoretical grounds,
they have to be considered on a par with the standard field theories as an immediate consequence of the existence of twin defects,
because for twin-like models their most relevant physical manifestations are completely indistinguishable.  This is the case, e.g., for effective field theories resulting from the integration of UV degrees of freedom, where higher kinetic terms are naturally induced.

\section*{Acknowledgement}
The authors acknowledge financial support from the Ministry of Science and Investigation, Spain (grant FPA2008-01177), 
the Xunta de Galicia (grant INCITE09.296.035PR and
Conselleria de Educacion), the
Spanish Consolider-Ingenio 2010 Programme CPAN (CSD2007-00042), and FEDER.

\end{document}